# Combining fiber Brillouin amplification with a repeater laser station for fiber-based optical frequency dissemination over 1400 km

**Sebastian Koke[1], Alexander Kuhl[1], Thomas Waterholter[1], Sebastian M.F. Raupach[1], Olivier Lopez[2], Etienne Cantin[2,3], Nicolas Quintin[2], Anne Amy-Klein[2], Paul-Eric Pottie[3], and Gesine Grosche[1]**

[1] Physikalisch-Technische Bundesanstalt, Bundesallee 100, 38116 Braunschweig, Germany
[2] Laboratoire de Physique des Lasers, Université Paris 13, Sorbonne Paris Cité, CNRS, Villetaneuse, France
[3] LNE-SYRTE, Observatoire de Paris, Université PSL, CNRS, Sorbonne Université, Paris, France

E-mail: sebastian.koke@ptb.de

## Abstract

We investigate optical frequency dissemination over a 1400 km long fiber link in looped configuration over a pair of underground fibers between Braunschweig and Strasbourg. This fiber link is the first to combine fiber Brillouin amplifiers with a repeater laser station. Phase-coherent operation over more than five days is demonstrated. We analyze the repeatability of the performance over four campaigns and present results of 65 days in total. The weighted mean of the fractional frequency offset of the transferred optical frequency over the complete data set is $(-1.1 \pm 0.4) \times 10^{-20}$. By analyzing the stabilization signals of the two individual fibers, the correlation of the phase noise on the two fibers is shown to be >98%.

## 1. Introduction

The fiber-based transfer of optical frequencies reaches uncertainties $\ll 10^{-18}$ over continental distances [1][2][3] and is currently the only means enabling comparison of remote state-of-the art optical clocks [4][5][6][7] without significant uncertainty contributions of the link [8]. Such a low remote clock comparison uncertainty is neither achievable through the alternative of microwave-based satellite connections [9] nor by exchanging transportable optical clocks [10][11]. Besides supporting the roadmap towards the redefinition of the SI-second [12], fiber-based comparisons of distant optical clocks enables high-resolution measurements of height differences by chronometric levelling [13][14][15] or tests of non-local relativistic effects [16].

Driven by, e.g., the effort to connect the clocks located at Europe's national metrology institutes, the length of the phase-stabilized fiber links [17] has been scaled up to continental distances of >1000 km [1][2][3][18]. One obstacle to be overcome for this scaling is the signal attenuation along the link: light propagating over 1000 km in optical fiber exhibits an attenuation of >200 dB. Uni-directional Erbium doped fiber amplifier (EDFA) are widely used for compensating these losses in fiber telecommunication networks. High-performance phase-coherent optical frequency dissemination requires bi-directional, common-path signal transmission along the complete fiber link [17]. Usage of bi-directional EDFA (bEDFA) variants turned out to be problematic for optical frequency dissemination over distances greater than a few hundreds of kilometers: due to back-reflections at connectors, unideal splices or Rayleigh scattering, the metrological signal on bEDFA-based fiber links tends to exhibit self-oscillations and in turn amplitude fluctuations [3][19]. These increase with the number of cascaded bEDFAs and eventually cause signal outages [3][19] leading to "cycle slips" (CSs) in the phase-locked loop (PLL) employed for optical path-length stabilization [20]. Such CSs are equivalent to an error of the transferred optical frequency. For an optical frequency around the minimum of fiber attenuation (≈200 THz), a CS rate of 1/100 s corresponds to a fractional frequency uncertainty of $5\times10^{-17}$. This is inadequate for the comparison of today's best optical clocks having a $10^{-18}$ uncertainty level. The negative impact of CSs in clock comparisons can be avoided by detecting the CSs [18][19][21] and disregarding the affected data points. This procedure, however, breaks with the phase-coherent averaging behavior of optical links, if no further measures ensuring phase coherence across the outages are applicable and implemented [3][22]. Inhibiting amplification-induced amplitude fluctuations and CSs in the first place, requires decreasing the modulation depth of the amplitude fluctuations by operating the bEDFAs at a lower gain of ≈15 dB [2]. Since equipment can only be installed into available huts along a fiber link and the huts have a typical spacing of 100 km, an excess loss of ≥50 dB is accumulated along a 1000 km fiber link. Hence, the signal-to-noise ratio (SNR) of the phase-noise-stabilization beat is lowered and the CS rate may rise due to a static but low SNR [20].

Repeating the signal by phase locking a repeater laser to the weak signal on the link [2][23][24] has been pursued as an alternative amplification technique to overcome this loss of SNR. As the repeater's output power is independent of the power of the received signal, excess loss is compensated, and amplitude modulations are not passed on. Over the last years, this repeater scheme has been incorporated into repeater laser stations (RLSs), which operate reliably and automatically in field [2][24]. In addition to repeating, these RLSs incorporate phase stabilizations allowing for segmenting a fiber link into individually stabilized shorter spans [25]. By such a segmenting, the travel-time-induced delay limit of phase noise cancellation [17] is improved allowing to cope with fiber spans exhibiting high phase-noise levels [25]. Moreover RLSs also provide a phase-coherent signal output for link performance evaluation, frequency comparison with another links' output signal (see below) or any local use. Using the combination of RLSs and bEDFAs on a 1100 km long four-span link, an optical link has been demonstrated on 3,5 days (with only 73 points removed, due to CSs) and a frequency dissemination uncertainty of $1.3\times10^{-19}$. Increasing the length to 1480-km led to the necessity for a thorough adjustment of the link parameters and frequency dissemination uncertainty of $9\times10^{-20}$ has been demonstrated after 12 hours (4 points removed) [2]. Reliable link operation required indeed careful adjustment of the bEDFA gain settings for preventing the onset of self-oscillations [2]. The bEDFAs should also be correctly engineered for their gain to be stable and not subject to temperature-induced variations [2][26].

An alternative means for bi-directional amplification is offered by either stimulated Raman scattering [27] or stimulated Brillouin scattering [1][28][29] in the transmission fiber itself. Fiber Brillouin amplifier (FBA) excels due to the narrow amplification bandwidth of ≈10 MHz and a gain of up to 50 dB [28]. As a result of the propagation direction sensitive gain and the saturation of the Brillouin gain with input power [28][31] beat signals obtained on FBA-based links have a high amplitude stability. Therefore, low CS rates are achievable. In 2015, an FBA-based 1400 km long fiber link between the Physikalisch-Technische Bundesanstalt (PTB) and the University of Strasbourg (UoS) and back was demonstrated with CS-free measurement up to a duration of ≈5 days (CS rate ≪ $10^{-5} s^{-1}$). The uncertainty of the transferred optical frequency was <$6\times10^{-20}$ for individual runs and <$2\times10^{-20}$ when averaged over multiple runs [1]. However, two limitations are apparent that may prevent similar performance for even longer distances. First, FBA-based fiber links are subject to spontaneous Brillouin backscattering background. For the average distance of ≈200 km between available installation points for the in-field FBA modules (FBAMs) on the fiber loop PTB–UoS–PTB, optical power at the input of the amplifier is only ≈100 nW. This resulted in an achievable SNR of the phase-noise-stabilization beat of only ≈30 dB (in 100 kHz resolution bandwidth) limited by the spontaneous Brillouin backscattering background, which makes link performance susceptible to polarization changes and amplifier gain variations. Second, the phase-noise-stabilization bandwidth of an L=1400 km long fiber link is only 35 Hz. Fiber links over even longer distances may therefore exhibit stabilized phase noise levels above the $L^{3/2}$-scaling based extrapolations due to the lack of suppression of the often observed bump in the free-running phase noise density around 20 Hz [1][3].

In this paper, we investigate the performance, i.e. the instability and the accuracy, of a link combining FBAs with an RLS for the first time. This link set-up has already been used successfully for several international clock comparisons with clock uncertainties down to $2\times10^{-17}$ [8][16][32]; we now give an in-depth study of the link's uncertainty limit, with a total amount of data equivalent to 65 days.

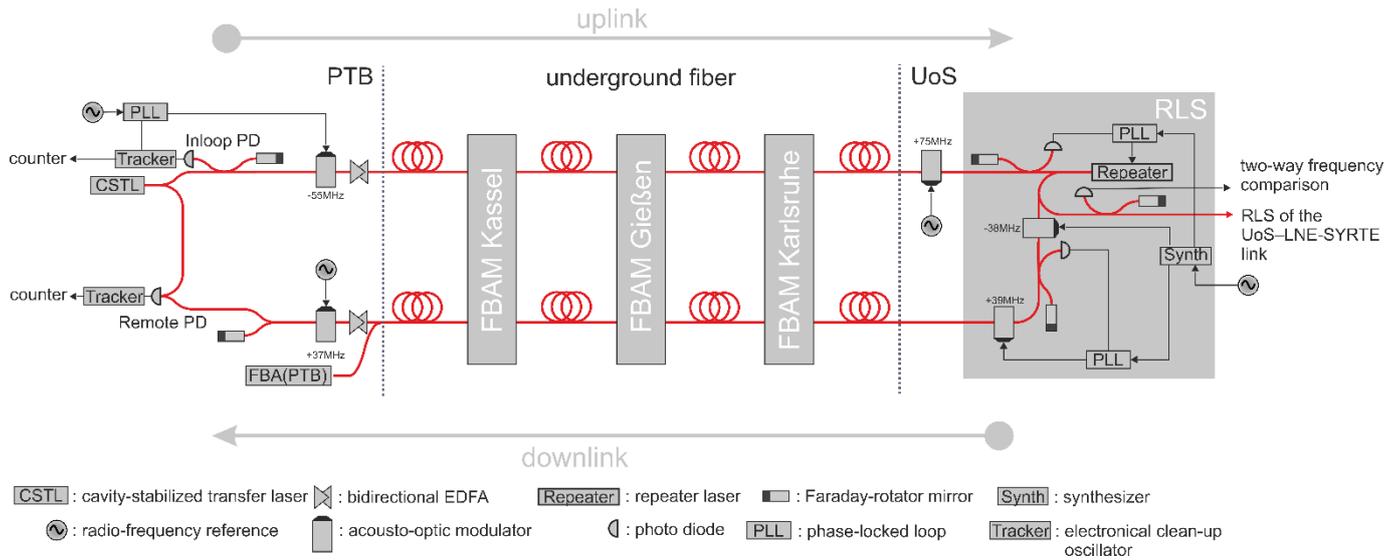

*Figure 1: Experimental setup of the fiber link PTB-UoS-PTB.*

In detail, we test the combination of three in-field FBAMs with one RLS installed at the University of Strasbourg (UoS) on the looped fiber link PTB–UoS–PTB (see Fig. 1 for details). The RLS segments the loop into an uplink PTB→UoS and a downlink UoS→PTB, each individually stabilized. We compare the performance of the cascade of up- and downlink to the performance of the single-span, FBA-based loop over the same fiber connection [1]. We also investigate the repeatability of the optical frequency transfer performance. To support the relevance of our findings for future comparisons of optical clocks with an estimated systematic uncertainty <10⁻¹⁸, we quantify an assumption which is the basis for the out-of-loop verification of the frequency transfer to UoS, namely the correlation of the phase noise on up- and downlink. At the end of this paper, we discuss the lay-out for a hypothetic, 3,000 km long fiber link based on the demonstrated performance of the combination of FBAs with RLSs.

## 2. Set-up of the fiber link

Fig. 1 shows the set-up of the fiber link. The link is formed by two telecommunication fibers within the same ≈700 km long fiber cable connecting PTB and the UoS. Previously, this fiber link has been operated as a single-span loop only using FBAs with the two fibers being patched together at UoS [1]. For performing clock comparisons [8][16][32], an RLS has been inserted at UoS. Light tapped off from this RLS and a second RLS serving the UoS–LNE-SYRTE link is used for performing two-way frequency comparison between the two links [8]. With the insertion of the RLS, the loop PTB–UoS–PTB is replaced by the cascade of an uplink, stabilized at PTB, and a downlink stabilized by the RLS at UoS.

As the metrological signal, we use a cavity-stabilized transfer laser near 194.4 THz. Between PTB and UoS, three FBAMs are installed in the computing centers of the University of Kassel, Justus Liebig University in Gießen, and the Karlsruhe Institute of Technology [1]. These FBAMs amplify light in both directions on both fibers. The pump lasers of the FBAMs are phase stabilized to the signal received from PTB on the uplink [29]. One additional FBA at PTB is amplifying the received downlink signal travelling from UoS to PTB. An additional frequency shift has been introduced in front of the RLS in order to operate the FBAs without hardware modifications at the same optical frequencies as in the case of the single-span loop.

For the characterization of the optical frequency dissemination over the cascade of uplink and downlink, we measure at PTB the deviation between the optical frequency received on the downlink and the uplink's source light frequency via the "remote" beat. In optical clock comparisons, this beat serves as an out-of-loop characterization of the frequency dissemination to UoS over the uplink. We later verify the underlying assumption of correlated phase noise on up- and downlink. To truthfully characterize a remote setting with independent temperature fluctuations of starting and end-point, the in-loop and remote interferometer are placed in two separate housings. However, they still reside close to each other in the same temperature-controlled laboratory.

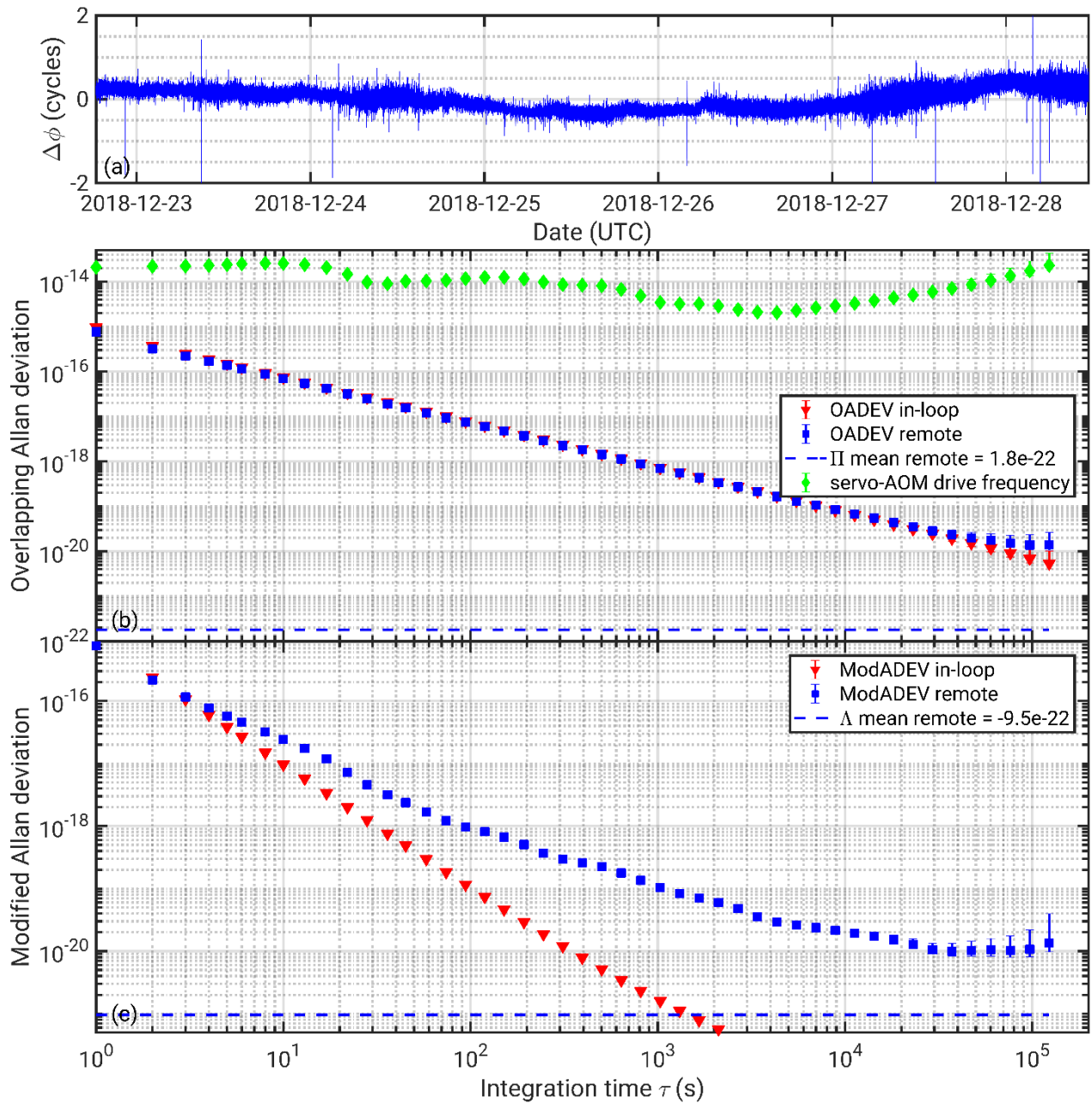

*Figure 2: Results of the longest continuous measurement run of all campaigns. (a) Times series of the phase deviation derived from Λ-counted 1 s data. (b) and (c) OADEV and modADEV of the in-loop beat (red), the remote beat (blue), and the uplink servo AOM's drive frequency (green).*

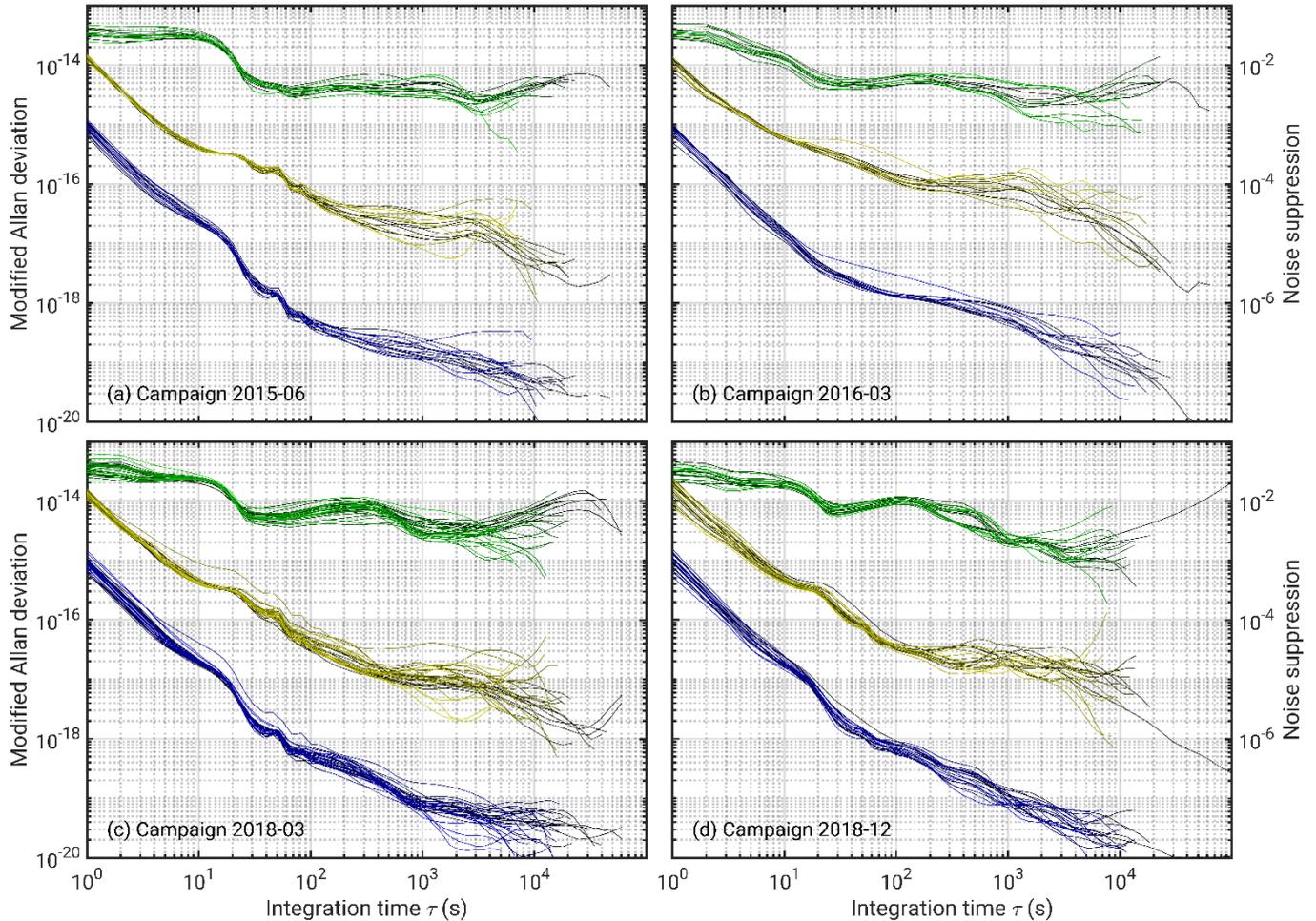

*Figure 3: Scatter of the modified Allan deviation of all >20ks long continuous segments: uplink servo AOM's drive frequency (green, left axis), remote (blue, left axis), and the corresponding noise suppression (yellow, right axis). Different runs are coded by the color's brightness.*

We typically achieve an SNR of ≈35–40 dB for the uplink in-loop stabilization beat and ≈30–35 dB for the remote beat, at an offset of ±1 MHz in a resolution bandwidth of 100 kHz. These SNRs are at or above the 30 dB that we found to be necessary for long-term reliable link stabilization. The difference in the achieved SNR is partly due to the design of the FBA modules currently installed in field, which do not permit a fully independent control of the frequency of the four pump waves (two fibers, two directions). The Brillouin frequencies of the distributed gain regions inside the buried fibers differ on the MHz-level and, hence, a compromise between all of them must be found. In this optimization, we give more weight to the uplink's signal levels to ensure a correct frequency dissemination to Strasbourg. The beats used for the repeater laser lock and the downlink stabilization typically have an SNR of ≈30 dB at an offset of ±1 MHz in a resolution bandwidth of 100 kHz.

For high-frequency phase noise rejection, electronical clean-up oscillators, called tracking filter, are phase locked to the beat notes at PTB with a loop bandwidth of ≈100 kHz.

Consistency of this clean-up step is checked by analyzing the difference between two of these clean-up oscillators, whose free-running frequency offset to the nominal beat frequency is of similar magnitude ($|\Delta f| \approx 500$ kHz) but of opposite sign. Using this scheme, we expect to detect malfunctions of this filtering step (cycle slips of the tracking PLLs), resulting from, e.g., glitches of the beat note amplitude, as frequency difference between the two clean-up oscillators [21].

The frequencies of the clean-up oscillators and the drive frequency of the servo AOM are frequency counted using a dead-time-free totalizing counter (K&K FXE80). Internally these counters operate with a gate time of 1 ms (Π counting up to this time span) and we set these counters to average the phase over 1000 samples to retrieve a Λ-counted sample over a 1 s long interval ($\Lambda_{1s}$) [1][30][33], which permits uncorrupted Fourier analysis and computation of modified Allan deviations.

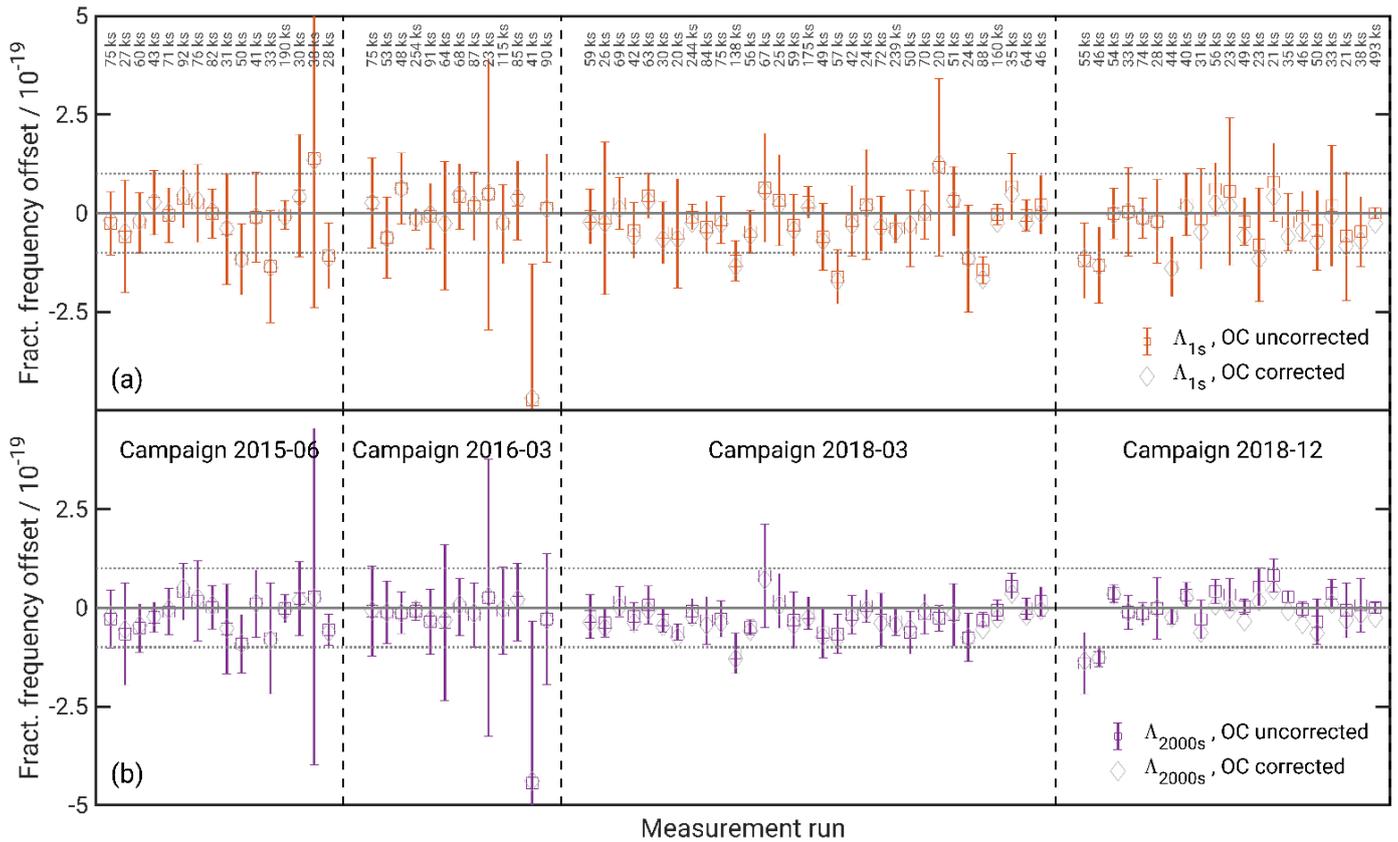

*Figure 4. Unweighted mean of fractional frequency offset of (a) $\Lambda_{1s}$ data and of (b) $\Lambda_{2000s}$ data (see text) of all measurement runs longer than 20 ks. The error bars are determined by the OADEV(N/4) value, N being the number of $\Lambda_{1s}$ and $\Lambda_{2000s}$ data points. While the colored data squares show the raw remote frequency offsets, the open grey triangles show the offsets corrected for the known overcompensation (OC) at the remote end.*

## 3. Results

Using this set-up, we have performed four link characterization campaigns each lasting over several weeks: the first one from 1. June to 24. June 2015 (parallel to the first international clock comparison campaign [8][16][32]); the second from 14. March to 30. April 2016; the third from 27. March 2018 to 2. May 2018; and the fourth from 3. December 2018 to 31. December 2018. In the following, we report the results over segments with continuous link operations covering more than 5.6 million data points in total. The link set-up has evolved during the time span of the assessment. For instance, on PTB side in 2017 the fiber connection between in-loop and remote interferometer has been shortened and link stabilization, tracking filters, and FBAs have been upgraded with auto-relocking capabilities. Furthermore, the SNR of the PTB beats has been enhanced by 5 dB. The RLS in Strasbourg was replaced by another one beginning of 2018.

For the standardized assessment of the frequency dissemination performance in these campaigns over the semi-automated fiber link, we track the in-loop and remote beat signal twice and analyze $\Lambda_{1s}$ data points where the tracking filter difference and the offset from the known beat frequencies has been below 10σ (σ being the corresponding $\Lambda_{1s}$ standard deviation), and also check for a working lock of the transfer laser to the ultra-stable cavity and of the pump laser of the FBA at PTB. In addition to that, we manually filter out time spans, where we observe intermittent frequency excursions on both remote tracking filter at PTB. This is the typical behavior we observe if the level and/or SNR of the PLLs signal drops. As PTB signals were tracked consistently and also the uplink in-loop tracking filter do not show increased noise at these time spans, we suspect a deterioration of the downlink stabilization signal at UoS as the cause for these excursions. In future, log data from the stabilization signal of the RLS could be used to confirm this assumption. Apart from enhancing the beat SNR in general, one further way of mitigation could be the upgrade of the FBAs with automatic polarization tracking, as dynamic polarization rotation might be the cause for the amplification degradation and in turn of the link stabilisation deterioration. These effects, however, do not limit the resolution achievable in optical clock comparisons over the fiber link, as filtering of these anomalies can be applied in clock comparisons [34]. In the following we will only consider measurement runs with >20 ks of continuous operation.

## 3.1 Frequency transfer uncertainty over the cascaded fiber link PTB–UoS–PTB

Figure 2 shows the performance of the longest continuous operation run within these campaigns spanning over 493 ks. Fig 2 (a) shows the time series of the phase deviation of the remote beat and shows a CS-free operation over more than 5 days.

The overlapping Allan deviation (OADEV) of the out-of-loop beat shown in Fig. 2 (b) starts off with $7.6 \times 10^{-16}$ at an integration time $\tau = 1$ s. The OADEV averages down as $1/\tau$ until $\tau \approx 10$ ks and reaches an instability of $1.4 \times 10^{-20}$ for an averaging time of $\tau = 123$ ks. The $\Pi$-averaged offset of the remote beat frequency is $1.8 \times 10^{-22}$ and shows that zero compatible operation is achieved in this run.

Comparing these results with the performance of the single-span FBA-loop requires a closer look at our link set-up. Musha et al. have discussed the instability reduction achievable by segmenting a fiber link for the case of uncorrelated but equal free-running fiber noise on the individual segments [25]. However, a looped layout over two fibers running in the same cable exhibits correlated noise contributions, as has already been discussed earlier [3][35]. In this case, the achievable delay-limited phase noise density [17] is a factor of four higher for the single-span loop than for the cascade of up- and downlink[1]. Hence, even in the presence of correlation on a looped link, the short-term instabilities are expected to halve when switching from a single-span link (subscript ssl) to the cascade of individually stabilized uplink (subscript u) and downlink (subscript d) in the case of constant fiber noise: $\sigma_y^{ud} = \sigma_y^{ssl}/2$.

For the single-span loop, a short-term instability of $\sigma_y^{ssl}(1s) = 2 \times 10^{-15}$ has been obtained [1] over a 320 ks long time series with an instability of the drive frequency servo acousto-optic modulator (AOM) of $\sigma_y^{ssl,AOM}(1 \; s) = 7 \times 10^{-14}$. We regard this drive frequency as an estimate for the free-running fiber-induced noise as it is the major contribution. Minor contributions arise from, e.g., the offsets from the radiofrequency references in Strasbourg and interferometer arm length variations. The AOM drive frequency instability values correspond to a noise suppression of $\sigma_y^{ssl}(1 \; s)/\sigma_y^{ssl,AOM}(1 \; s) = 2.9 \times 10^{-2}$ at 1 s integration time. This value is nearly the expected factor of two higher than the corresponding noise suppression for the cascaded up- and downlink of $\sigma_y^{ud}(1 \; s)/[2 \; \sigma_y^{u,AOM}(1 \; s)] = 1.8 \times 10^{-2}$. The residual discrepancy from the factor of two is within the variations we observe for different runs (see later).

The modified Allan deviation (modADEV) of the remote signal shown in Fig. 2(c) initially averages down slightly faster than for white-phase noise, i.e. approximately proportional to $\tau^{-1.7}$ [18][36]. For longer averaging times the slope gradually decreases. The curve can be approximated by: $\text{mod } \sigma_y^{ud}(\tau) = \left[ \left( 6.0 \times 10^{-16} \; \tau^{-3/2} \right)^2 + (9.0 \times 10^{-17} \; \tau^{-1})^2 + \left( 1.2 \times 10^{-18} \; \tau^{-1/2} \right)^2 + (1.0 \times 10^{-20})^2 \right]^{1/2}$. At an averaging time of 123 ks, the modADEV reaches a floor of $1.3 \times 10^{-20}$.

In Figure 3 we analyze the repeatability of the instabilities and show the variations of the modADEV curves of the uplink servo AOM's drive frequency and the remote beat over all continuous runs longer than 20 ks. Similar to earlier findings [9][37], we observe a difference in the fiber phase noise between day and night and between weekday and weekend, which is probably induced by human activity. As a result, the short-term instability of fiber-induced frequency noise and the short-term instability of the remote beat frequency scatter. For the June 2015 campaign, for instance, the 1 s modADEV values of the uplink servo AOM's frequency fluctuation and of the remote beat frequency spread both by a factor of two from $2.4 \times 10^{-14}$ to $4.6 \times 10^{-14}$ and from $6.2 \times 10^{-16}$ to $1.2 \times 10^{-15}$, respectively. The in-loop beat exhibits the same spread (not shown). A similar spread by a factor of 2–3 in the short-term modADEVs is also observable in the other campaigns. We observe typical fractional corrections in the low to mid $10^{-14}$ range applied by the servo loop to the servo AOM's drive frequency, with values reaching up to $2 \times 10^{-13}$ for eight runs in the Dec. 2018 campaign. For three of the four campaigns, the remote modADEVs show a bump around 15 s and at 50 s, where the magnitude of the latter is about an order of magnitude below the former. Such bumps are explainable by a harmonic disturbance with a period of 35 s, which may hint

---

[1] As shown in Fig. 5, the phase noise on the two fibers forming the uplink and the downlink is correlated. Hence, the phase noise at the point forming the remote end in this characterization is given by $4 \; S_{fiber}(f)$, with $S_{fiber}(f)$ being the one-way phase noise on each of the both fibers. This impacts the single-span loop and the cascade of up- and downlink in the same way. The delay-limited phase noise of a single-span loop at the remote end is given by $S_D^{ssl}(f) \approx a \left( \frac{2\pi f}{c_n} \right)^2 (2L)^2 \; 4 \; S_{fiber}(f)$. For the cascade of up-

and downlink the stabilization delay halves leading to the delay-limited phase noise of $S_D^{ud}(f) \approx a \left( \frac{2\pi f}{c_n} \right)^2 L^2 \; 4 \; S_{fiber}(f)$. The corresponding $\Lambda_{1s}$ instabilities are obtained via integration [39]: $\sigma_D^x(f) = \left[ \int_0^\infty S_D^x(f) \; |W(f)|^2 \; df \right]^{1/2}$, with $W(f)$ being the spectral weight associated with the counting process [30][39].

at a temperature stabilization as the possible origin. Interestingly, the set of modADEV curves are ordered from 1 s to typically an averaging time of ≈10 s. This may indicate a generic time scale of stationary noise conditions on our fiber link.

The noise suppression mod $\sigma_y^{u,d}(\tau)/[2\,\mathrm{mod}\,\sigma_y^{u,fiber}(\tau)]$ is shown in the yellow traces. For purely white phase noise, one expects a scaling of the noise suppression with $\tau^{-3/2}$ [30], which is also the approximate slope of the shown curves up to averaging times of ≈10 s. At 1 s, the noise suppression values scatter around a value of ≈$1.3\times10^{-2}$: as already discussed, this agrees with the expected factor of two improvement when compared to the single-span loop. For the December 2018 campaign, the short-term noise suppression scatters significantly more than for the other campaigns. This may result from the heavy construction works taking place at that time in the computing center of Kassel University. These works also explain why the mean length of continuous operation in this campaign is on the order of half a day. The spread of the noise suppression curves widens for all campaigns above an averaging time of ≈100 s, typically accompanied by a decrease of the slope of the noise suppression with averaging time. As the modADEV-levels are of similar magnitude as for the single-span loop at these averaging times, we believe that the same mechanism of residual differential drift of in-loop and remote interferometer poses the limitation here. Such a drift may depend on the level of environmental perturbations during the measurements and may therefore explain the larger spread of the noise suppression.

The accuracy achieved for the FBA-RLS-based fiber link is summarized in Figure 4 for $\Lambda_{1s}$ data and $\Lambda_{2000s}$ data and >20 ks long runs ($\Lambda_{2000s}$ data are obtained from the available $\Lambda_{1s}$ by numerically extending the $\Lambda$-weighting to 2000 s [1]). All fractional frequency offsets scatter in the low $10^{-20}$ range with only four out of the 82 runs having a $\Lambda_{2000s}$ fractional frequency offset above $\pm1\times10^{-19}$. The scatter of the $\Lambda_{2000s}$ fractional frequency offset of the FBA-RLS combination are larger than for the single-span loop [1], which may indicate the presence of additional out-of-loop processes in the cascaded configuration, due to for instance the RLSs interferometer. The observed offsets are mostly zero-compatible within 1 σ, with σ being the statistical uncertainty of the average [39]. The lower statistical uncertainties in the 2018's campaigns reflect the improvement in the instabilities already observable in Fig. 3.

The $1/\sigma^2$-weighted mean of the $\Lambda_{2000s}$ fractional frequency offsets of the four campaigns are $(-2.1\pm1.6)\times10^{-20}$, $(-1.0\pm1.7)\times10^{-20}$, $(-2.5\pm0.6)\times10^{-20}$, and $(2.5\pm5.7)\times10^{-21}$. The $1/\sigma^2$-weighted mean of all campaigns is $(-1.1\pm0.4)\times10^{-20}$. These values in the low $10^{-20}$ range are comparable to the weighted frequency offset achieved for the single-span loop.

Different processes may induce these residual offsets. The mismatch between frequency of the unsteered hydrogen maser acting as radio-frequency reference at PTB and the GPS-derived radio-frequency reference at Strasbourg can not explain this shift due to the specific engineering of RLS, which cancels any offset or instabilities of its local RF oscillator [2][24]. For the single-span loop, the overcompensation due to the slight optical frequency mismatch in the forward and backward propagation direction [35] was found to match quite well with the observed offsets [1]. Here, however, correcting for this overcompensation increases the observed offsets and the weighted averages (see grey open symbols in Fig. 4)². Furthermore, the offsets may originate from uncompensated paths between the two interferometers at PTB and inside the RLS and differential drift of their short arms. For the typical length of the measurement runs and the campaigns, however, we expect these contributions to average out. In addition, polarization mode dispersion [17] may cause offsets.

## 3.2 Correlation analysis of the free-running phase noise of uplink and downlink fiber

The independent stabilization of uplink and downlink in the FBA-RLS-based fiber loop allows studying the above-mentioned correlations of the phase noise on the two fibers, which is the basis for the out-of-loop verification of the frequency transfer to UoS in optical clock comparisons. During the double passage, light returning to PTB at the uplink (subscript u) and light returning to UoS at the downlink (subscript d) has accumulated the phase noise given by

$$\Phi_u(t) = \int_0^L \delta\varphi_u(t - 2\tau + z/c_n, z)\,dz$$
$$+ \int_0^L \delta\varphi_u(t - \tau + z/c_n, L - z)\,dz$$

and

$$\Phi_d(t) = \int_0^L \delta\varphi_d\left(t - 2\tau + \frac{z}{c_n}, L - z\right)dz$$
$$+ \int_0^L \delta\varphi_d\left(t - \tau + \frac{z}{c_n}, z\right)dz,$$

respectively. Here $\delta\varphi_x(z,t)$ denotes phase noise on fiber $x$ at time $t$ and position $z$, $\tau$ is the one-way travel time through the fiber and $c_n$ denotes the light's phase velocity inside the fiber. In locked operation, the PLLs apply a correction $\Delta f_u(t)$

---

² Correcting for the overcompensation is impacted by the unmonitored frequency offset between the chip-scale atomic clock (CSAC), used as reference for the drive frequency

synthesis of the AOM in front of the RLS, and the hydrogen maser used as radio-frequency reference at PTB. Based on the aging specifications of the CSAC, we estimate this impact to be minor on a level $<10^{-20}$.

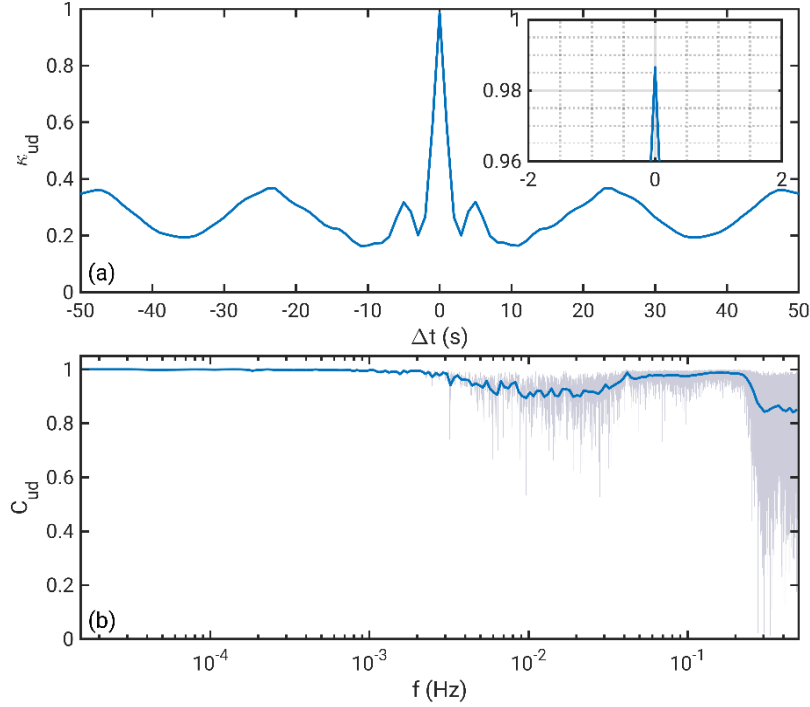

*Figure 5: Correlation of the $\Lambda_{ls}$ frequency corrections of the servo AOMs of up- and downlink. (a) Temporal correlation coefficient $\kappa_{ud}(\Delta t)$. The inset shows a zoom into the central peak showing a peak correlation coefficient of >0.98. (b) Coherence function $C_{ud}(f)$ (for definitions see text). The blue curve is the average of the grey curve on a logarithmically spaced grid.*

and $\Delta f_d(t)$ to the servo AOMs on up- and downlink, respectively, to compensate for the accumulated phase noise. Figure 5 shows the temporal correlation of the frequency corrections $\Delta f_u(t)$ and $\Delta f_d(t)$ applied to the servo AOMs by the phase stabilization of uplink and downlink, respectively. For simultaneously recorded data, the correlation coefficient:

$$\kappa_{ud}(\Delta t) = \langle \Delta f_u(t) \Delta f_d(t + \Delta t) \rangle$$

reaches a value of greater than 0.98 at $\Delta t = 0$, indicating a strong correlation. A spectral analysis using the coherence function:

$$C_{ud}(f) = \frac{|S_{ud}(f)|^2}{S_{uu}(f) S_{dd}(f)}$$

reveals a strong correlation ($C_{ud}(f) \approx 1$) for frequencies below 0.002 Hz and between 0.04 and 0.2 Hz, which are well within the delay limited phase stabilizations bandwidth of ≈70 Hz ($S_{uu}(f)$, $S_{dd}(f)$): spectral density of up- and downlink frequency corrections, respectively; $S_{ud}(f)$: cross-spectral density). There are two Fourier frequency regions where a degradation of $C_{ud}(f)$ is observable: between 0.002 and 0.04 Hz and above 0.2 Hz. The former region corresponds to averaging times just above the characteristic kink we typically observe in the modADEV curves of the uplink correction signal at ≈20 s, as shown in Fig. 3. This kink appears at the same averaging times as the lower-frequency bump in the remote modADEV that has been attributed to a harmonic disturbance above. Such pronounced kinks are not observable in the modADEV of the single-span loop [1]. We therefore think that this degradation in the coherence function between the two correction signals is due to environmental perturbations at UoS, most probably the air climatisation, affecting the equipement installed in Strasbourg. The origin of the degradation above 0.2 Hz remains elusive.

As proved by above comparisons between the instability of the cascade of up- and downlink and the single-span loop instability in stabilized operation, the correlated free-running fiber noise translates into correlated residual noise on up- and downlink for sampling times of ≥1 s. Hence, the phase noise on up- and downlink adds constructively at these sampling times. This is different from the case of predominantly anticorrelated phase noise on the two fibers, for which phase noise contributions cancel out to a high degree for every sample, and the out-of-loop frequency instability may fall below the single fiber instability. This is also different from the case of uncorrelated phase noise on the two fibers, for which one may have partial compensation of fiber phase noise for single samples, while the out-of-loop phase noise induced instability is the quadratic sum of the single fiber noise [25]. Thus, the correlated phase noise on the pair of fibers in our setup allows us to assess the maximum transferred frequency error for every single sample using the out-of-loop round-trip signal, which is the most stringent evaluation possible.

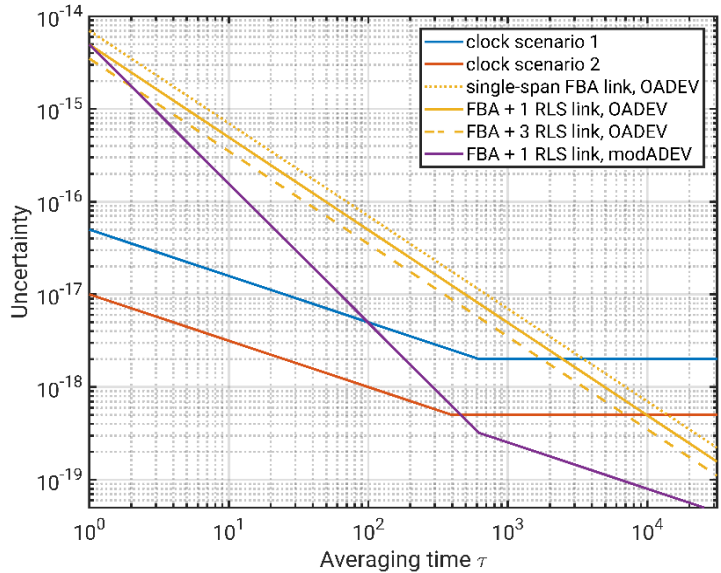

*Figure 6: Link and clock uncertainty for the discussed scenarios of a clock comparison over a 3000 km non-looped fiber link using $\Lambda_{I3}$ counting.*

## 4. Scaling fiber links beyond the demonstrated 1400 km

In closing, let us now reconsider the obtained results in view of scaling the fiber link length to a hypothetic distance of 3000 km. This is a realistic fiber length for, e.g., connections Berlin–Madrid, Boulder–Gaithersburg, or Beijing–Hong Kong. For such a non-looped 3000 km long fiber link, the 1 s instability is estimated to be $\approx 7 \times 10^{-15}$, given the results of the 1400 km single-span loop and assuming the applicability of L$^{3/2}$-scaling [17].

Let us further assume, the purpose of this fiber link is the comparison of optical clocks with the combination of top-performance instability of OADEV($\tau$)$\approx$5×10$^{-17}$ ($\tau$/s)$^{-1/2}$ [38] and a top-performance systematic uncertainty of $\approx$2×10$^{-18}$ [5][7] (blue curve in Fig. 6). As shown in Fig. 6, the OADEV of a single-span link would drop below the systematic uncertainty of such clocks after $\approx$3500 s. These long averaging times can be shortened by either extending $\Lambda$-averaging to longer times for a better fiber phase noise rejection [1][39] or by segmenting the link. Sub-dividing the link in two individually stabilized segments will enhance the locking bandwidth from 17 Hz to 33 Hz, and, hence, the often observed prominent phase noise contribution around 20 Hz [1][3] will still be suppressed. Including one mid-way segmentation, parity with the clock's systematic uncertainty is achieved for $\Pi$-counting after $\approx$2500 s. More importantly, however, simultaneous segmentation and repetition using an RLS mid-way will enhance beat SNRs by stripping off the

(amplified) spontaneous Brillouin scattering background. Using such a fiber link design, SNRs above the 30 dB achieved for the 1400 km single-span loop are feasible securing a reliable CS-free link operation.

For comparing clocks having improved performance of, e.g., an instability of OADEV($\tau$)$\approx$1×10$^{-17}$($\tau$/s)$^{-1/2}$ and a systematic uncertainty of 5×10$^{-19}$ (orange curve in Fig. 6), parity with the clock's instability and systematic uncertainty is only achieved >7000 s, even if the link is segmented into 4 spans (3 RLS). Resolution at the level of such improved clock's can be achieved significantly faster by continuing with phase-averaging up to 500 s ($\Lambda_{500s}$ counting), see violet curve in Figure 6. Thus the fiber link uncertainty contribution is reduced according to the modADEV [1][39], and with just one RLS, it falls below the clock's instability already at 500 s.

## 5. Conclusions

The presented results demonstrate that FBAs can be successfully combined with RLSs for the fiber-based dissemination of optical frequencies. Only four out of the 82 runs showed a $\Lambda_{2000s}$ fractional frequency offset exceeding 1×10$^{-19}$. This shows, that this combination supports frequency dissemination with an uncertainty better than one tenth of the estimated systematic uncertainty of today's best optical clocks [4][5][6][7]. These results provide a new technological option for laying out fiber links spanning even greater distances: using amplitude-stable, CS-free frequency dissemination based on FBA on fiber segments with moderate phase noise and employing RLS for improving phase-noise rejection and the SNR.

In addition, a correlation >98% has been found for the phase noise on the two fibers connecting PTB and UoS for a sampling time of 1 s. Thus, the frequency offset of the round-trip signal gives an upper limit for the frequency dissemination error to UoS for every 1 s-sample, which corroborates the validity of out-of-loop verification for optical clock comparisons.

These findings will help to perform optical clock comparisons reliably not only for time and frequency metrology but also for the prospect of addressing fundamental science [16] or as a long-haul chronometric levelling method for geodesy [13][14][15].

## Acknowledgements

We acknowledge H. Schnatz's long-standing support, T. Legero and U. Sterr for operating the cavity-stabilized laser, A. Koczwara for technical assistance, and E. Benkler for fruitful discussions. Furthermore, we thank the computing center teams for the enduring assistance. This work would not have been possible without the support of the GIP RENATER, and especially T. Bono and E. Camisard. We also acknowledge support from staff of computing centers of University of Strasbourg, University of Kassel, Justus Liebig University Gießen, and the Karlsruhe Institute of Technology.

This work has been funded by the European Metrology Programme for Innovation and Research (EMPIR) in project 15SIB05 (OFTEN) and 18SIB06 (TiFOON). The EMPIR is jointly funded by the EMPIR participating countries within EURAMET and the European Union. PTB thanks the DFG for financial support in the Sonderforschungsbereich 1128 Relativistic Geodesy and Gravimetry with Quantum Sensors, geo-Q under project A04 and in the Exzellenzcluster 2123 "QuantumFrontiers". LNE-SYRTE and LPL acknowledge support from the Agence Nationale de la Recherche (ANR blanc LIOM 2011-BS04-009-01, Labex First-TF ANR 10 LABX 48 01, Equipex REFIMEVE + ANR-11-EQPX-0039).